\documentclass[aps,showpacs,preprint]{revtex4}

\usepackage{amssymb}
\usepackage{latexsym}
\usepackage{amsfonts}
\usepackage{graphicx}
\usepackage{epsfig}
\usepackage{amsmath}
\usepackage{float}
\usepackage{verbatim}

\begin{document}

\title{Financial factor influence on scaling and memory of trading volume in stock market}

\author{Wei Li,$^1$ Fengzhong Wang,$^1$ Shlomo Havlin,$^{1,2}$ and
  H. Eugene Stanley$^1$}

\affiliation{$^1$Center for Polymer Studies and Department of Physics,
  Boston University, Boston, MA 02215 USA\\
  $^2$Department of Physics, Bar-Ilan University,
  Ramat-Gan 52900, Israel}

\date{\today}

\begin{abstract}
  
We study the daily trading volume volatility of 17,197 stocks in the
U.S. stock markets during the period 1989--2008 and analyze the time
return intervals $\tau$ between volume volatilities above a given
threshold $q$. For different thresholds $q$, the probability density
function $P_q(\tau)$ scales with mean interval $\langle\tau\rangle$ as
$P_q(\tau)=\langle\tau\rangle^{-1}f(\tau/\langle\tau\rangle)$ and the
tails of the scaling function can be well approximated by a power-law
$f(x)\sim x^{-\gamma}$. We also study the relation between the form of
the distribution function $P_q(\tau)$ and several financial factors:
stock lifetime, market capitalization, volume, and trading value. We
find a systematic tendency of $P_q(\tau)$ associated with these factors,
suggesting a multi-scaling feature in the volume return intervals. We
analyze the conditional probability $P_q(\tau|\tau_0)$ for $\tau$
following a certain interval $\tau_0$, and find that $P_q(\tau|\tau_0)$
depends on $\tau_0$ such that immediately following a short/long return
interval a second short/long return interval tends to occur. We also
find indications that there is a long-term correlation in the daily
volume volatility.  We compare our results to those found earlier for
price volatility.

\end{abstract}


\maketitle

\section{Introduction}
  
Because the dynamics of financial markets are of great importance in
economics and econophysics
\cite{Mandelbrot63,Mantegna95,Kondor99,Mantegna00,Takayasu97,Bouchaud00,Johnson03,Liu99,Plerou01},
the dynamics of both stock price and trading volume have been studied
for decades as a prerequisite to developing effective investment
strategies. Econophysics research has found that the distribution of
stock price returns exhibits power-law tails and that the price
volatility time series has long-term power-law correlations
\cite{Engle93,Ord85,Harris86,Pfleiderer88,Schwert,Pictet93,Pagan96,Ding96,Liu97,Cont98,Cizeau97,Pasquini99}.
To better understand these scaling features and correlations, Yamasaki
{\em et al.} \cite{Yamasaki05} and Wang {\em et al.}
\cite{Wang06,Wang07} studied the behavior of price return intervals
$\tau$ between volatilities occurring above a given threshold $q$. For
both daily and intraday financial records, they found that (i) the
distribution of the scaled price interval $\tau/\langle\tau\rangle$ can
be approximated by a stretched exponential function, and (ii) the
sequence of the price return intervals has a long term memory related to
the original volatility sequence. The scaling and memory properties of
financial records are similar to those found in climate and earthquake
data \cite{Bunde04,Bunde05,Livina05,Lennartz08,Altmann05,Lux96}.

A feature of the recent history of the stock market has been large price
movements associated with high volume. In the Black Monday stock market
crash of 1987, the Dow Jones Industrials Average (DJIA) plummeted 508
points, losing 22.6 percent of its value in one day, which led to the
pathological situation in which the bid price for a stock actually
exceeded the ask price. In this financial crash approximately
$6\times10^8$ shares traded, a one-day trading volume three times that
of the entire week previous. Understanding the precise relationship
between price and volume fluctuations has thus been a topic of great
interest in recent research \cite{Plerou00,Gopikrishnan00}. Trading
volume data in itself contains much information about market dynamics,
e.g., the distribution of the daily traded volume displays power-law
tails with an exponent within the L\'evy stable domain
\cite{Plerou07,Plerou09}.  Recently, Ren and Zhou \cite{Ren10} studied
the intraday database of two composite indices and 20 individual indices
in the Chinese stock markets. They found that the intraday volume
recurrence intervals show a power-law scaling, short-term correlations
and long-term correlations in each stock index.

In this study we analyze U.S. stock market data over a range broad
enough to allow us to identify how several financial factors
significantly affect scaling properties. We study the daily trading
volume volatility return intervals $\tau$ between two successive volume
volatilities above a certain threshold $q$, and find a range of
power-law distributions broader than that found earlier in price
volatility return intervals \cite{Yamasaki05,Wang06}. We find a unique
scaling of the probability density function (PDF) $P_q(\tau)$ for
different thresholds $q$. We also perform a detailed analysis of the
relation between volume volatility return intervals and four financial
stock factors: (i) stock lifetime, (ii) market capitalization, (iii)
average trading volume, and (iv) average trading value. We find
systematically different power-law exponents for $P_q(\tau)$ when
binning stocks according to these four financial factors. Similar to
that found for the Chinese market \cite{Ren10}, we find that in the
U.S. stock market the conditional probability distribution,
$P_q(\tau|\tau_0)$ for $\tau$ following a certain interval $\tau_0$,
demonstrates that volume return intervals are short-term correlated. We
also find that the daily volume volatility shows a stronger long-term
correlation for sequences of longer lifetime but no clear changes in
long-term correlations for different stock size factors such as
capitalization, volume, and trading value.

\section{Data}

In order to obtain a sufficiently long time series, we analyze the daily
trading volume volatility of 17,197 stocks listed in the U.S. stock
market for at least 350 days. We obtain our data from the Center for
Research in Security Prices (CRSP) US stock database, which lists the
daily prices of all listed NYSE, Amex, and NASDAQ common stocks, along
with basic market indices. The period we study extends from 1 January
1989 to 31 December 2008, a total of 5042 trading days.

\section{Distribution of Volume Volatility Return Intervals}
  
For a stock trading volume time series, in a manner similar to stock
price analysis \cite{Cizeau97,Pasquini99,Wang06}, 
we define two basic measures: volume return $R$ and
volume volatility $\nu$. The volume return $R$ is defined as the
logarithmic change in the successive daily trading volume for each
stock, 
\begin{equation}
R(t)\equiv\ln\left(\frac{V(t)}{V(t-1)}\right),
\label{return.eq}
\end{equation}
where $V(t)$ is the daily trading volume at time $t$. We define volume
volatility to be the absolute value of the volume return. In order to
compare different stocks, we determine the volume volatility $\nu(t)$ by
dividing the absolute returns $|R(t)|$ by their standard deviation,
\begin{equation}
\nu(t)\equiv\ \frac{|R(t)|}{(\langle|R(t)|^2
\rangle-{\langle|R(t)|\rangle}^2)^{1/2}}, 
\label{volatility.eq}
\end{equation}
where $\langle\cdots\rangle$ is the time average for each stock. The
threshold $q$ is thus measured in units of standard deviation of
absolute volume return $|R(t)|$.

\begin{figure*}
\begin{center}
   \includegraphics[width=0.65\textwidth, angle = -90]{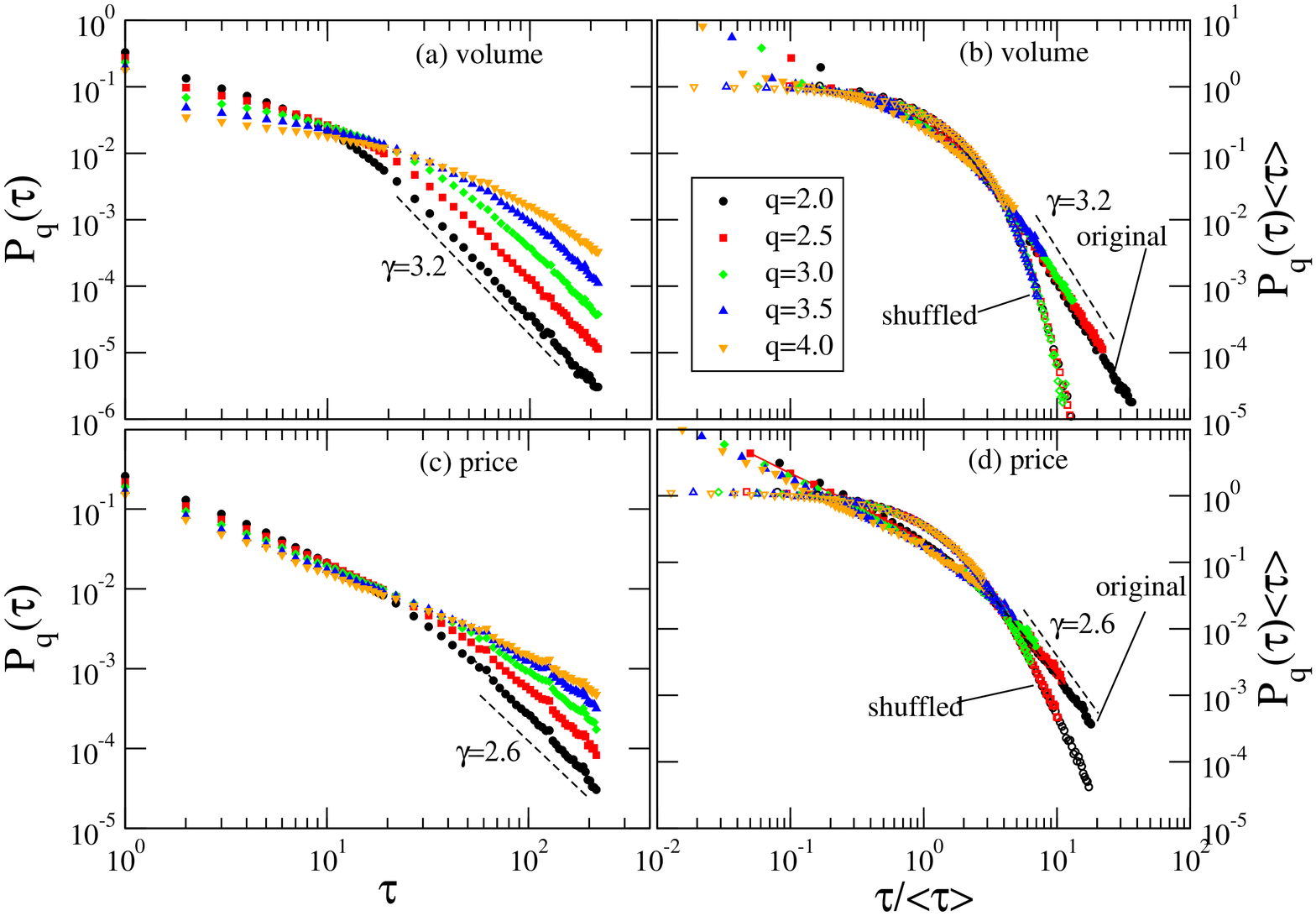}
\end{center}
\caption{(Color online) Probability distributions of volume volatility return intervals and price volatility return intervals for 17197 stocks. 
  Full symbols with different shapes represent 
  different thresholds $q$ varying from 2.0
  to 4.0. (a) Distribution of volume volatility return intervals, $P_q(\tau)$ versus $\tau$. (b) Scaled distribution of volume
  return intervals (full symbols) $P_q(\tau)\langle\tau\rangle$
  versus $\tau/\langle\tau\rangle$, and distribution of volume return
  intervals for shuffled volatility records (open symbols). The four
  curves with full symbols collapse onto one single curve, indicating
  a universal scaling function. The tail of scaling function is
  approximately power-law distribution, $f(x)\sim x^{-\gamma}$, with
  $\gamma\cong3.2$, while the curve fitting the shuffled records is
  exponential function, $f(x)=e^{-ax}$, from Possion distribution. A Poisson distribution indicates no correlation in shuffled volatility data, but the original dataset suggests strong correlation in the volatilities. The power-law exponents for intraday volume recurrence intervals of several Chinese stock indices are from $\gamma=1.71$ to $\gamma=3.27$ \cite{Ren10}. For comparison, (c) and (d) show the distribution and scaled distribution of price volatility return intervals respectively. Note the narrow range of power-law compared to (a).}
\label{Fig1}
\end{figure*}

For a volume volatility time series, we collect the time intervals
$\tau$ between consecutive volatilities $\nu(t)$ above a chosen
threshold $q$ and construct a new time series of volume return intervals
$\{\tau(q)\}$. Fig.~\ref{Fig1}(a) shows the dependence of $P_q(\tau)$ on $q$,
where $P_q(\tau)$ is the PDF of the volume volatility return interval
time series $\{\tau(q)\}$. Obviously, $P_q(\tau)$ decays more slowly for
large $q$ than for small $q$. For large $q$, $P_q(\tau)$ has a higher
probability of having large interval values because extreme events are
rare in a high threshold series. We next determine whether there is
there any scaling in the distribution by plotting the PDFs of the volume
return intervals $P_q(\tau)$, scaled with the mean volume return
interval $\langle\tau(q)\rangle$, for different thresholds in
Fig.~\ref{Fig1}(b). We can see that all five threshold values $q$ curves (full
symbols) callapse onto a single curve, suggesting the existence of a
scaling relation,
\begin{equation}
P_q(\tau)=\frac{1}{\langle\tau\rangle}f
\left(\frac{\tau}{\langle\tau\rangle}\right).
\label{pdf.eq}
\end{equation}
As the threshold $q$ increases, the curve (rare events) tends to be
truncated due to the limited size of the dataset. The tails of the
scaling function can be approximated by a power-law function as shown by
the dashed line in Fig.~\ref{Fig1}(b),
\begin{equation}
f\left(\frac{\tau}{\langle\tau\rangle}\right)\sim
{\left(\frac{\tau}{\langle\tau\rangle}\right)}^{-\gamma}, 
\label{scaling.eq}
\end{equation}
where the tail exponent is $\gamma$. The exponent of the scaled PDFs for
$q=2$ is $\gamma\cong3.2$ by the least square method, which is the same
as the unscaled PDF exponent $\gamma\cong3.2$ as shown in Fig.~\ref{Fig1}(a). The
power-law exponents for intraday volume recurrence intervals of several
Chinese stock indices are from $\gamma=1.71$ to $\gamma=3.27$
\cite{Ren10}. Our exponents $\gamma$ are larger than those in the
Chinese stock markets. This might be due to differing definitions of
volume volatility. In Ref.~\cite{Ren10}, the volume volatility is
defined as intraday volume divided by the average volume at one specific
minute of the trading day averaged over all trading days. Here we
define the volume volatility to be the logarithmic change in the
successive daily volumes [Eqs.~\ref{return.eq} and ~\ref{volatility.eq}]. For comparison, and using
the same approach, Fig.~\ref{Fig1}(c) and Fig.~\ref{Fig1}(d) show the analogous results
for price volatilities (see also the studies in
Refs.~\cite{Yamasaki05,Wang06,Jung08}).  Note that it is not easy to
distinguish between a stretched exponential and a power-law when
studying price volatilities \cite{Yamasaki05}, i.e., the power-law range
is small and a stretched exponential could also provide a good fit. In
contrast, the PDFs of the volume volatility return intervals display a
wide range of power-law tails, which differs from the stretched
exponential tail apparent in the price return intervals
\cite{Wang06}. Our results for volume volatility may suggest that
$P_q(\tau)$ for price volatility is also a power-law, but this could not
be verified because the range of the observed power-law regime [see
  Figs.~\ref{Fig1}(c) and~\ref{Fig1}(d)] is more limited than the broad range of scales
seen in the volume volatility [Figs.~\ref{Fig1}(a) and~\ref{Fig1}(b)]. The difference
between the power-law and stretched exponential behavior of $P_q(\tau)$
may be related to the existence or non-existence respectively of
non-linearity represented in the multifractality of the time
series. When non-linear correlations appear in a time record, Bugachev
{\em et al.} \cite{Bogachev07} showed that $P_q(\tau)$ is a power-law.
On the other hand, when non-linear correlations do not exist and only
linear correlation exists, Bunde {\em et al.} \cite{Bunde05} found
stretched exponential behavior.

A comparison with the shuffled records allows us to see how the
empirical records differ from randomized records. We shuffle the volume
volatility time series to make a new uncorrelated sequence of
volatility, and then collect the time intervals above a given threshold
$q$ to obtain synthetic random control records. The curve that fits the
shuffled records [the open symbols in Fig.~\ref{Fig1}(b)] is an exponential
function, $f(x)=e^{-ax}$, and forms a Possion distribution.  A Poisson
distribution indicates no correlation in shuffled volatility data, but
the empirical records suggest strong correlations in the volatility.

\section{financial factors}

\begin{figure*}
\begin{center}
   \includegraphics[width=0.65\textwidth, angle = -90]{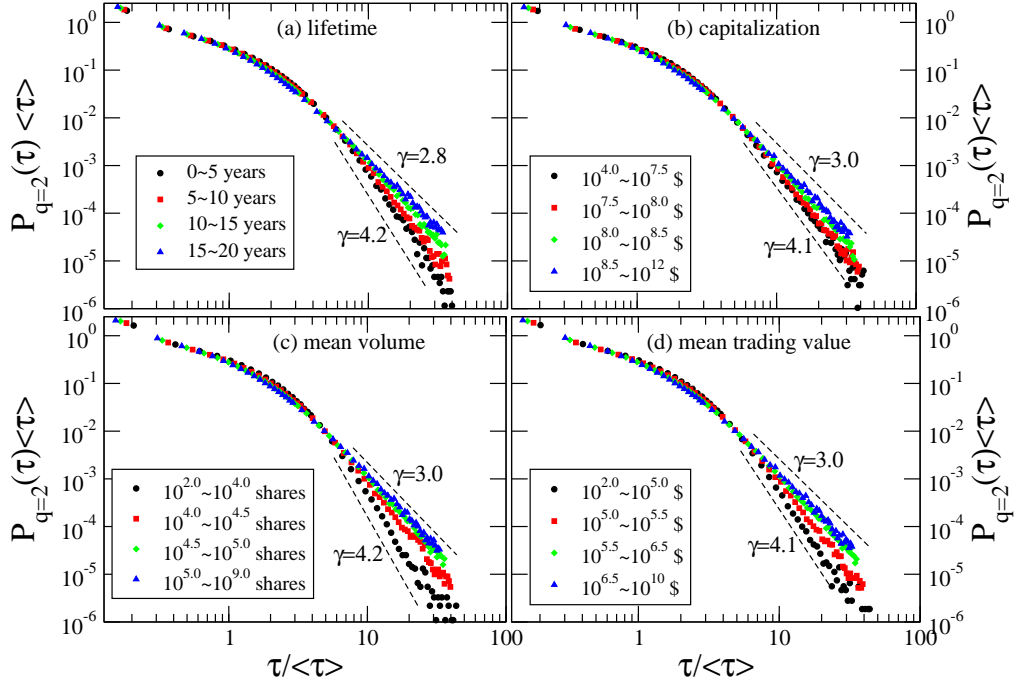}
\end{center}
\caption{(Color online) Relations between distribution function
  $P_2(\tau)\langle\tau\rangle$ of volume volatility return intervals and four
  financial factors: (a) lifetime, (b) market capitalization, (c)
  average daily trading volume, (d) average daily trading value, for the
  threshold $q=2.0$. The distrition functions decay with various exponents
  $\gamma$ and show similar systematic tendency for four financial factors.}
\label{Fig2}
\end{figure*}

We study the relation between the scaled PDFs
$P_q(\tau)\langle\tau\rangle$ as a function of $\tau/\langle\tau\rangle$
for four financial factors: (a) stock lifetime, (b) market
capitalization, (c) mean volume, and (d) mean trading value for
threshold $q=2.0$. For higher $q$ values, we do not have sufficient data
for conclusive results \cite{Bogachev07}. In Fig.~\ref{Fig2}, we plot the scaled
PDFs for these four factors. The volume return intervals characterize
the distribution of large volume movements.  A high probability of
having a large volume return interval $\tau$ suggests a correlation in
volume volatility, because small volatilities are followed by small
volatilities and the time interval between the two large volatilities
becomes relatively longer than those of random records. In order to
charaterize how these four factors affect the distribution of volume
return intervals, we divide all stocks into four bins for each
factor. In Fig.~\ref{Fig2}(a), the probability that $\tau$ will be large is
greater in the bin with 15$\sim$20 year old stocks (triangles) than in
the bins of younger stock. This indicates that small volatilities (below
the threshold) tend to follow small volatilities and that the time
intervals between large volatilities in the bin of 15$\sim$20 year-old
stocks are larger than the time intervals in the bin of 5 years old
stocks (dots). This also suggests that the volume volatility time
records of older stocks are more auto-correlated than those of younger
stocks.  The decaying parameters represented by the power-law exponents
are quite different: $\gamma\cong4.2$ for the shortest lifetime bin and
$\gamma\cong2.8$ for the longest lifetime bin. This significant
difference might be caused by differences in autocorrelation in these
series.
 
\begin{figure*}
\begin{center}
   \includegraphics[width=0.65\textwidth, angle = -90]{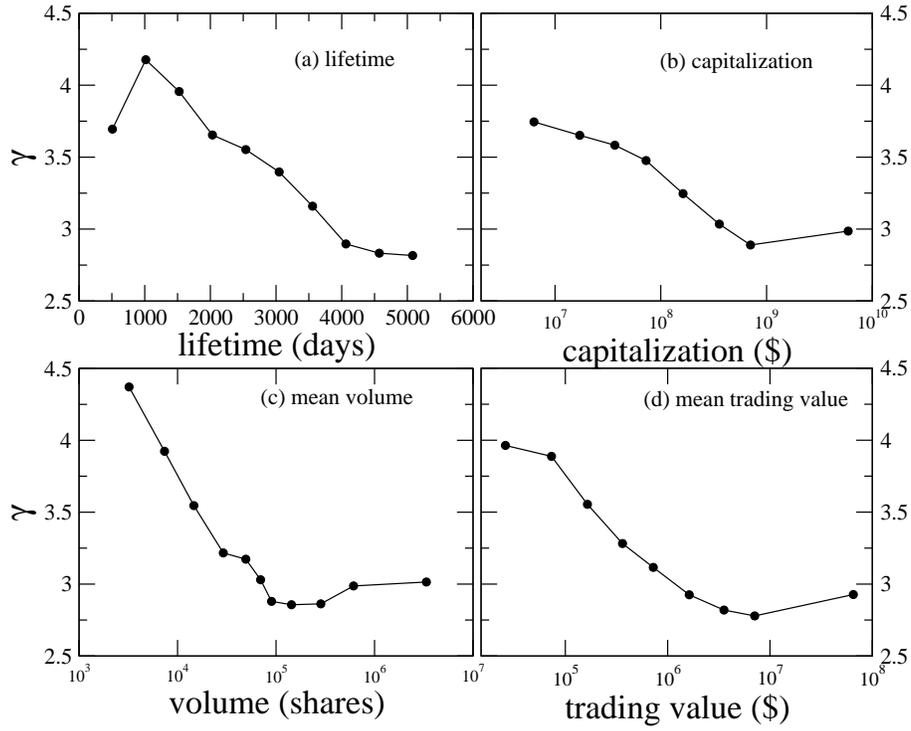}
\end{center}
\caption{ The power-law tail exponent $\gamma$ for different subsets of stocks. (a) Stocks are sorted into 10 subsets of different lifetime. Exponent $\gamma$ are abtained by fitting the PDF of volume volatility return intervals for each subset; (b) stocks are sorted into 8 subsets for different capitalization; (c) stocks are sorted into 11 subsets for different mean volume; (d) stocks are sorted into 9 subsets for different trading value.}
\label{Fig3}
\end{figure*}

In Figs.~\ref{Fig2}(b),~\ref{Fig2}(c), and~\ref{Fig2}(d), we show a similar tendencies for stock
bins with different capitalizations, mean volumes, and mean trading
values. Trading value is defined as stock price multiplied by
transaction volume. For each stock, we designate the lifetime average of
capitalization, volume, and trading value as performance indices. For
example, the power-law exponents of the PDFs,
$P_q(\tau)\langle\tau\rangle$, increase as the capitalization becomes
larger [see Fig.~\ref{Fig2}(b)]. To clarify the picture, we divide all stocks
into different subsets and study the behavior of the power-law exponent
$\gamma$ with regard to these four factors. In Fig.~\ref{Fig3}(a), stocks are
sorted into 10 subsets, from 508 days (2 years) to 5080 days (10
years). We fit the power-law tails of the volume return intervals for
each subset and plot the exponent $\gamma$ versus the lifetime of the
stocks. In Fig.~\ref{Fig3}(a), we can observe a systematic trend with stock lifetime. 
It is seen that a smaller exponent $\gamma$ which indicates a stronger
correlation in older stock subsets. Similarly, we sort the stocks by
capitalization, mean volume, and mean trading value, as shown in
Figs.~\ref{Fig3}(b),~\ref{Fig3}(c), and~\ref{Fig3}(d). It is seen that $\gamma$ decreases with increasing
of all these three factors but seem to become constant for large values of 
capitalizations, mean volumes and mean trading values. 

Since all factors similarly affect the scaling of the PDF,
$P_q(\tau)\langle\tau\rangle$, we now determine how much these factors
are correlated. To study the relations between different stock bins, we
plot the relation between trading value versus capitalization, mean
volume versus capitalization, and mean trading value versus mean volume
for all the stocks shown in Fig.~\ref{Fig3}. We see that larger capitalization
stocks tend to have a larger trading volume and a larger trading value,
which is consistent with Figs.~\ref{Fig1}(b),~\ref{Fig1}(c), and~\ref{Fig1}(d). The correlation
coefficients between trading value and capitalization, mean volume and
capitalization, and trading value and volume are 0.62, and 0.55, and 0.78,
respectively. The correlation coefficients are high because these
capitalization, volume, and trading value factors are all affected by
firm size. Our analyses do not, however, show a significant relationship
between stock lifetime and its trading value, capitalization, and mean
volume, and the correlation coefficients are all $<0.20$.

\section{Short-term Memory Effects}
  
We characterize a sequence of volume return intervals in terms of the
autocorrelations in the time series. If the volume return intervals
series are uncorrelated and independent of each other, their sequences
can be determined only by the probability distribution. On the other
hand, if the series is auto-correlated, the preceding value will have a
memory effect on the values following in the sequence of volume
volatility return intervals.

In order to investigate whether short-term memory is present, we study
the conditional PDF, $P_q(\tau|\tau_0)$, which is the probability of
finding a volume return interval $\tau$ immediately after an interval of
size $\tau_0$. In records without memory, $P_q(\tau|\tau_0)$ should be
identical to $P_q(\tau)$ and independent of $\tau_0$. Otherwise,
$P_q(\tau|\tau_0)$ should depend on $\tau_0$. Because the statistics for
$\tau_0$ of a single stock are of poor quality, we study
$P_q(\tau|\tau_0)$ for a range of $\tau_0/\langle\tau\rangle$. The
entire dataset is partitioned into eight equal-sized subsets,
$Q_1,Q_2,...Q_8$, with intervals of increasing size
$\tau_0/\tau$. Figure~\ref{Fig5} shows the PDFs $P_q(\tau|\tau_0)$ for $Q_2$,
i.e., small interval size $0.2<\tau_0/\langle\tau\rangle<0.4$ and $Q_6$
large interval size $3.2<\tau_0/\langle\tau\rangle<6.4$ for different
$q$. The probability of finding large $\tau/\langle\tau\rangle$ is
larger in $Q_6$ (open symbols) than in $Q_2$ (full symbols), while the
probability of finding small $\tau/\langle\tau\rangle$ is larger in
$Q_2$ than that in $Q_6$. Thus large $\tau_0$ tends to be followed by
large $\tau$, and vice versa, which indicates short-term memory in the
volume return intervals sequence. Moreover, note that $P_q(\tau|\tau_0)$
in the same subset for different thresholds $q$ fall onto a single
curve, which indicates the existence of a unique scaling for the
conditional PDFs as well. Similar results were found for the volume
volatility of the Chinese markets \cite{Ren10} and for price
volatilities \cite{Yamasaki05, Wang06}.

\section{Long-term Memory Effects}

In previous studies, the price volatility series was shown to have
long-term correlations. Using a similar approach, we test whether the
volume volatility sequence also possesses long-term correlations. To
answer this question, we employ the detrended fluctuation analysis (DFA)
method \cite{Peng94,Bunde00,Kantelhardt01} to further reveal memory
effects in the volume volatility series. Using the DFA method, we divide
an integrated time series into boxes of equal length $n$ and fit a least
squares line in each box. Next we compute the root-mean-square
fluctuation $F(n)$ of the detrended time series within a window of $n$
points and determine the correlation exponent $\alpha$ from the scaling
function $F(n)\sim n^\alpha$, where $\alpha\in [0,1]$. The correlation
exponent $\alpha$ characterizes the autocorrelation in the sequence. The
time series has a long-term memory and a positive correlation if the
exponent factor $\alpha>0.5$, indicating that large values tend to
follow large values and small values tend to follow small values. The
time series is uncorrelated if $\alpha=0.5$ and anti-correlated if
$\alpha<0.5$.

Using the DFA method, we analyze the price volatility and volume
volatility time series by plotting in bins the relation between
correlation exponent $\alpha$ and the four financial factors, including
stock lifetime, market capitalization, mean trading volume, and mean
trading value.  All the price volatility and volume volatility
correlation exponents are significantly larger than 0.5, suggesting the
presence of long-term memory in both price volatility sequences and
volume volatility sequences. In all of the plots, the price volatility
series shows a stronger long-term correlation than the volume volatility
series. Moreover, as shown in Fig.~\ref{Fig6}(a), $\alpha$ on average increases
for the stocks with a lifetime ranging from 350 days to 3800 days (about
15 years), and then shows a slight decrease, suggesting that
long-lasting stocks tend to have a persistent price and volume movement
on large scales. The increasing exponent $\alpha$ indicates that the
volume volatility of older stocks is more correlated than that of
younger stocks. This is consistent with the indication in Fig.~\ref{Fig2}(a) that
the volume volatility of older stocks are more auto-correlated.
Figures~\ref{Fig6}(b), ~\ref{Fig6}(c), and ~\ref{Fig6}(d) show that there is no systematic tendency
relation between $\alpha$ and market capitalization, trading volume, and
trading value.

\section{conclusions}

We have shown the scaling properties and memory effect of volume
volatility return intervals in large stock records of the
U.S. market. The scaled distribution of volume volatility return
intervals displays unique power-law tails for different thresholds
$q$. We also find different power-law exponents $\gamma$ of $P_q(\tau)$
for the four essential stock factors: stock lifetime, market
capitalization, average trading volume, and average trading value. These
different exponents may be related to long-term correlations in the
interval series. Significantly, the daily volume volatility exhibits
long-term correlations, similar to that found for price volatility. The
conditional probability, $P_q(\tau|\tau_0)$ for $\tau$ following a
certain interval $\tau_0$, indicates that volume return intervals are
short-term correlated.

\newpage

\begin{figure*}
\begin{center}
   \includegraphics[width=0.65\textwidth, angle = -90]{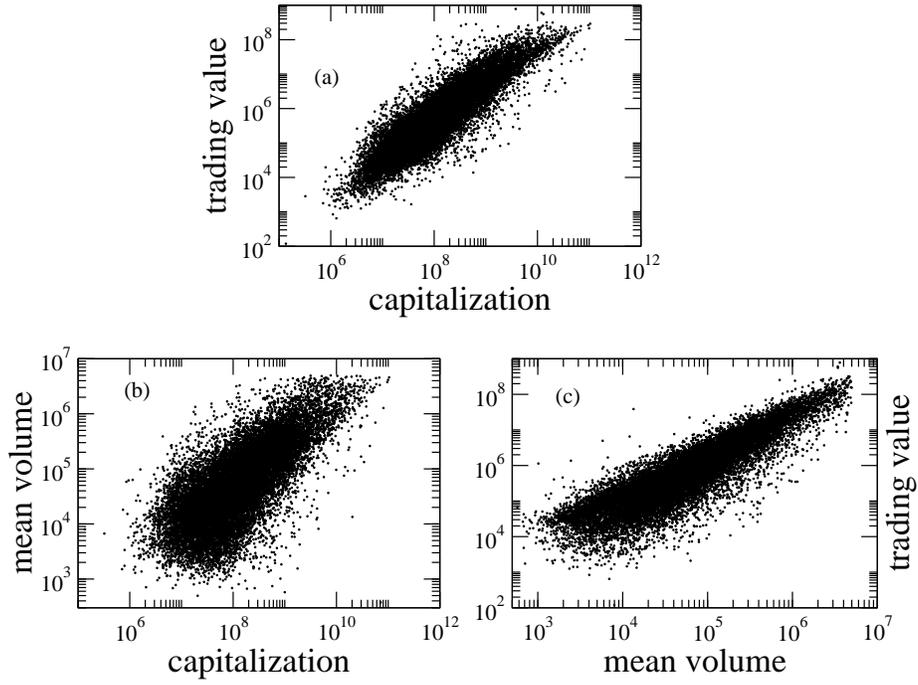}
\end{center}
\caption{Scatter plots for the relations in stocks 
between trading value and capitalization, mean volume and capitalization, 
trading value and mean volume for all 17,197 stocks. 
For example, a point on panel (a) represents a stock, which has 
\$  $10^8$  capitalization and \$ $10^6$ average trading value. The correlation coefficients between trading value and capitalization, mean volume and capitalization, trading value and volume are 0.62, and 0.55, and 0.78 respectively.}
\label{Fig4}
\end{figure*}

\begin{figure*}
\begin{center}
   \includegraphics[width=0.65\textwidth, angle = -90]{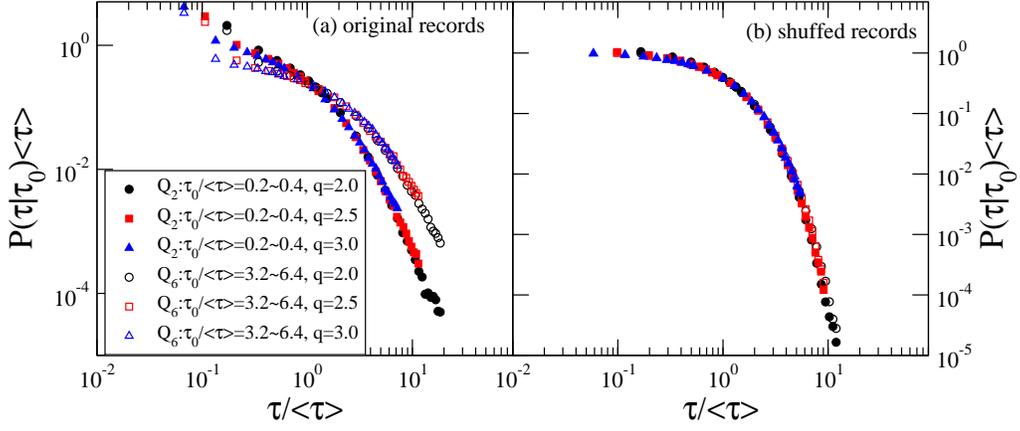}
\end{center}
\caption{(Color online) Conditional PDF $P_q(\tau|\tau_0)$ of volume volatility return intervals $\tau$ for different thresholds $q=2.0, 2.5, 3.0$, as a function of $\tau/\langle\tau\rangle$ for different $\tau_0/\langle\tau\rangle$ bins. A small $\tau_0$ subset $Q_2$ (full symbols) and a large $\tau_0$ subset $Q_6$ (open symbols) are displayed in (a). For example, subset $Q_6$ contains events of finding $\tau$ after large interval $3.2<\tau_0/\langle\tau\rangle<6.4$. In contrast to subset $Q_6$, subset $Q_2$ has larger probability to be followed
  by small $\tau/\langle\tau\rangle$ and smaller probability to be
  followed by large $\tau/\langle\tau\rangle$, which indicates short
  term correlation in the records: small intervals are followed by
  small intervals and large intervals are followed by large
  intervals. There is no memory effect in shuffle records as seen in (b) that the PDFs of all the subsets collapse onto one curve.}
\label{Fig5}
\end{figure*}

\begin{figure*}
\begin{center}
   \includegraphics[width=0.65\textwidth, angle = -90]{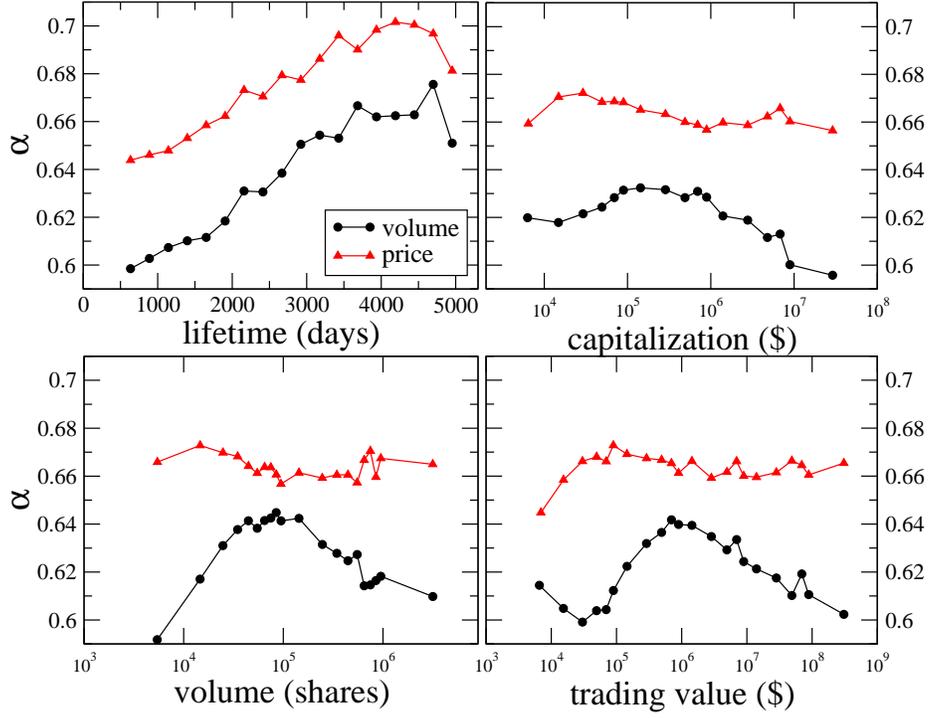}
\end{center}
\caption{(Color online) Correlation exponent $\alpha$ obtained from
  detrended fluctuation analysis (DFA) of volume volatility (square) and
  price volatility (triangle). The plot shows the relation between
  $\alpha$ and four factors: (a) lifetime, (b) market capitalization,
  (c) average daily trading volume, (d) average daily trading value, for
  the threshold $q=2.0$.}
\label{Fig6}
\end{figure*}

\end{document}